\documentclass{article}

\usepackage{nejlt}
\usepackage{amsmath}
\usepackage{graphicx}
\usepackage{physics}
\usepackage{qcircuit}
\usepackage{lipsum}

\title{Discussion of Quantum Consensus Algorithms}
\author{Samuel Fulton}

\begin{document}

\abstract{ Leader election is a crucial process in many areas such as cloud computing, distributed systems, task orchestration, and blockchain. Oftentimes, in a distributed system, the network needs to choose a leader, which would be responsible for synchronization between different processors, data storage, information distribution, and more. In the case where the network is anonymous, no classical algorithms could solve the problem exactly. However, in the setting of quantum computers, this problem is readily solved. In this paper, we analyze the quantum consensus algorithm developed by Seiichiro Tani. We look at the inner workings of the algorithm and develop a circuit representation of the key steps. We review Mochon's fault tolerant leader election algorithm. We then implement a simple leader election algorithm on a quantum computer.
}

\maketitle
 \section*{Introduction}
In this paper, we will introduce distributed systems and consensus algorithms. We then zoom in and look at a specific class of distributed systems known as anonymous or symmetric distributed systems. From there, we introduce the classical approach to breaking symmetry and electing a leader. We show that there is no classical deterministic algorithm for leader election among an anonymous distributed system. We then introduce two quantum algorithms developed by Seiichiro Tani in  [\newcite{Tani}] for anonymous leader election. The quantum algorithms for anonymous leader election are more than computational speedups; they are deterministic solutions to a classically non-deterministic problem. We implement the second quantum leader election algorithm in Qiskit. Lastly, we analyze work by Mochon and  Kitaev, [\newcite{Mochon}], on developing fault-tolerant leader election algorithms.

 \section*{Distributed Systems}  
We will start with a brief definition of consensus and distributed systems. We define a distributed system to be a collection of communicating processors all working towards a common goal. Let’s consider a toy example. Imagine a distributed system of computers attempting to factor a large number $N$. Once one processor “believes” that it has factored the number, it proposes its prime factors, $P_i$, of N. The other processors in the distributed system check if $\prod_i P_i=N$. If the majority of the processors agree that  $\prod_i P_i=N$ then $P_i$ is accepted as the factors of $N$. All of the processors can now move on to factoring a different prime number $M$, all the while remembering that the prime factors of $N$ have been decided. Other examples where consensus algorithms come into play include leader election, blockchain, load balancing, clock synchronization, and more. 

 Three things characterize a distributed system: Agreement, Validity, and Termination. As the name suggests agreement means that all non-faulty processors must agree on the same value. In the case of factoring $N$, all non-faulty processors must either agree that $P_i$ are the factors of N or all agree that $P_i$ are not the factors of $N$. Validity is the assertion that under non-Byzantine conditions, the distributed system will never return an incorrect result. In the example of the processors attempting the factor $N$, this would mean that given sufficient time all non-faulty processors would find the same prime factors of $N$. Termination is the assertion that given enough time the processors are guaranteed to complete the task. \\
\indent What do we mean when by non-faulty processors?  There are two types of faulty processors. The first is a processor that experienced a crash. This processor stops responding to other processors in the distributed system. This is a common occurrence. However, since distributed systems only need the majority of processors to work, crash failures are readily dealt with. The second type of faulty processor experiences Byzantine failure. Byzantine failure occurs when a processor malfunctions in a way such that it sends incorrect data to the distributed system. An example of a Byzantine processor is a hacked processor. For consensus problems having Byzantine failures is the worst scenario. Fault-tolerant consensus algorithms address crash and Byzantine failure. The most widely used consensus algorithms in distributed and cloud computing systems are Paxos and its variants such as Raft. These algorithms are used for leader election and typically tolerate non-Byzantine failures. 

 \section*{Quantum Consensus} 
 Mazzarella categorizes quantum consensus into four classes in [\newcite{Mazzarella}]. $\sigma$-expectation consensus, reduced state consensus, symmetric state consensus, and single $\sigma$-measurement consensus. Consider a quantum network consist of three qubits, and three observables of the form.
\begin{align*}
    \sigma^1 = \sigma^z \otimes I \otimes I,\\
    \sigma^2 = I\otimes \sigma^z \otimes I,\\
    \sigma^3 = I \otimes I \otimes \sigma^z.
\end{align*}
The system is in consensus concerning the expectation of $\sigma z$
if 
\begin{align*}
 Tr(\rho\sigma^1) = Tr(\rho\sigma^2) = Tr(\rho\sigma^3).
\end{align*}
Noted that Quantum consensus is achieved by a quantum network rather than traditional computational resources. They are similar counterparts, though we must take account of probabilistic outcomes due to the stochastic nature of quantum mechanics. We will show how quantum entanglements can offer an advantage in terms of reaching an agreement in a distributed setting.
 \section*{Leader Election}
As described in [\newcite{Brooker}], "leader election is the simple idea of giving one thing (a process, host, thread, object, or human) in a distributed system some special powers. Those special powers could include the ability to assign work, the ability to modify a piece of data, or even the responsibility of handling all requests in the system." Leader election is extremely useful for improving efficiency. A leader can often bypass consensus algorithms and simply inform the system about changes that will be made. Leaders can help with consistency because they can see all of the changes that have been made to the system. By acting as a central data cache, a leader can improve consistency across the entire system.

A single leader does introduce some drawbacks. Namely, a single leader is a critical point for failure. If the leader crashes the entire distributed system may halt. Furthermore, if a single leader experiences Byzantine failure, the entire system may waste time following incorrect protocols. However, many of these drawbacks are mitigated through the use of consensus algorithms. Oftentimes, the improved efficiency of leader election out ways any drawbacks. In the next section, we will explore how leaders can be fairly elected.

\section*{Leader Election Algorithm I} 
One interesting consensus problem is the anonymous leader election. Anonymous leader election is used in the case where we have a collection of identical processors and wish to designate a leader. Figure 1 depicts an anonymous distributed system. In the system, processors do not have unique identification and all run the same protocol. Thus, the system is symmetric under all permutations.
\begin{figure}
\begin{align*}
\includegraphics[width=0.4\textwidth]{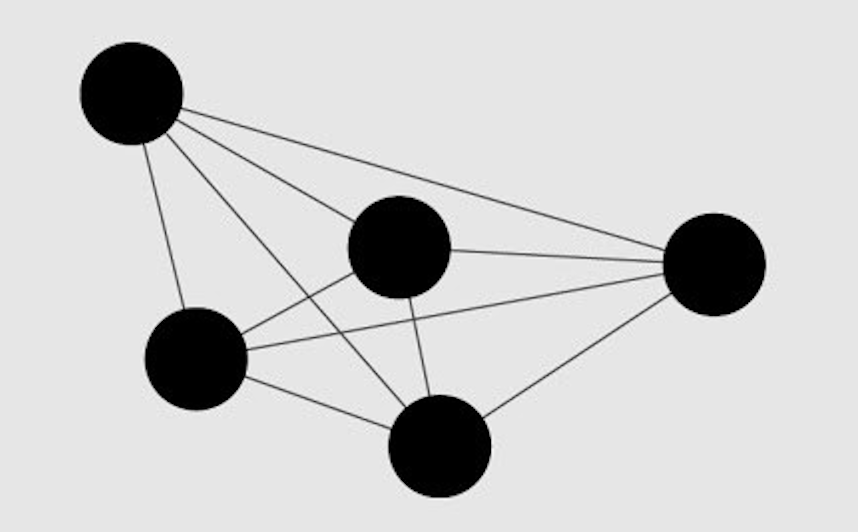}
\end{align*}
\caption{}
\end{figure}
  The symmetry of the system prevents non-probabilistic leader election. The classical approach outlined by Seiichiro Tani in [\newcite{STani}] is to install a coin flip in each processor. Each processor flips a coin if heads it is eligible for leader election. If the coin is tails, it is a follower. If multiple processors get heads, then the protocol is repeated with the eligible candidates. Note that this process is non-deterministic and has an expected run time of $O(\log(n))$, where $n$ is the number of processors. Let's compare this to a simplistic quantum leader election algorithm, which was proposed in [\newcite{Tani}]. We will start with a two-processor scenario. For this algorithm, we have two processors $A$ and $B$. We generate the quantum state
 
\begin{equation*}
    W = \frac{1}{\sqrt{2}}(\ket{01}+\ket{10}).
\end{equation*}
We send the first qubit to processor $A$ and the second qubit to processor $B$. If processor $A$ measures $\ket{1}$, we know processor $B$ must measure $0$ and vice versa. Whichever processor measures $\ket{1}$ becomes the leader. For $n$ processors we generate the state
\begin{align*}
    W_n &= \frac{1}{\sqrt{n}}(\ket{10...0} +  ... +\ket{0...01}),\\
        &=\frac{1}{\sqrt{n}} \sum_{k=0}^{n-1} \ket{2^k}.
\end{align*}
Each processor receives a qubit. Whichever processor measures $\ket{1}$ is the leader. This algorithm terminates after one run, which is in stark contrast to the non-deterministic classical leader election. 

 \section*{Algorithm II} 
 There is a more robust approach to quantum leader election proposed in [\newcite{Tani}]. Consider the case with processors $A$ and $B$. Each processor prepares the state
\begin{equation*}
    \ket{R_1}=\ket{R_2} = \frac{1}{\sqrt{2}}(\ket{0}+\ket{1}).
\end{equation*}
We send $\ket{R_1}$ and $\ket{R_2}$ through the following circuit, where $X$ is the Pauli matrix $X$ corresponding to a bit flip, and the remaining two gates are control not gates. 
\begin{figure}
 \begin{align*}
\includegraphics[width=0.45\textwidth]{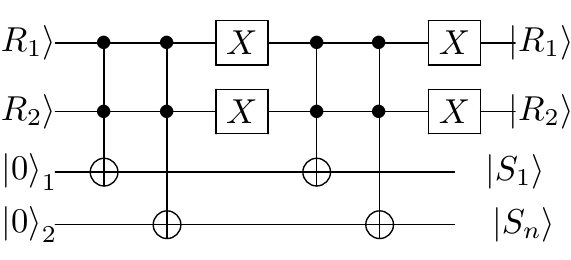}
\end{align*}
\caption{}
\end{figure}

The output of the circuit is the state
\begin{align*}
    \ket{R_1\: R_2 \: S_1 \: S_2} &= (\ket{00}+\ket{11})\ket{11} \\
    &+ (\ket{01}+\ket{10})\ket{00}.
\end{align*}
In words, $S_1$ and $S_2$ and both $\ket{1}$ if $\ket{R_1}$ and $\ket{R_2}$ are equal. Processor $A$ has the state $\ket{R_1\:S_1}$ and processor $B$ has the $\ket{R_2\:S_2}$ qubit. We note that $\ket{S_1}$ and $\ket{S_2}$ are entangled. After each processor measures its $S$ state, the system will collapse to either 
\begin{equation*}
    (\ket{00}+\ket{11})\ket{11}
\end{equation*}
or
\begin{equation*}
    (\ket{01}+\ket{10})\ket{00}.
\end{equation*}
If the system collapses to the $(\ket{01}+\ket{10})\ket{00}$ state, processor $A$ measures its $\ket{R_1}$ state and processor $B$ measures $\ket{R_2}$ state. Whichever processor measures $\ket{1}$ is the leader.
If the system collapses into the state $(\ket{00}+\ket{11})\ket{11}$, both processors apply the unitary operation 

\begin{equation*}
    U = \frac{1}{\sqrt{2}} 
    \begin{pmatrix}
    1 & -i \\
    -i & 1
    \end{pmatrix}
\end{equation*}
 
to their $S$ states. 
\begin{align*}
    &U_A \otimes U_B (\ket{00} + \ket{11}) \\
    &= U_A \ket{0} \otimes U_B \ket{0} + U_A \ket{1} \otimes U_B \ket{1}, \\
    &= (\ket{0} - i\ket{1})\otimes (\ket{0}-i\ket{1}) \\
    &+ (\ket{0} - i\ket{1})\otimes (\ket{0}-i\ket{1}),\\
    &= \ket{00}-i\ket{01}+i^2\ket{11}\\
    &+i^2\ket{00}-i\ket{01}-i\ket{10}+\ket{11},\\
    &= -i (\ket{01}+\ket{10}).
\end{align*}
Just like in the first case, processor $A$ measures its $\ket{R_1}$ state, and processor $B$ measures $\ket{R_2}$ state. Whichever processor measures $\ket{1}$ is the leader. 

This algorithm is readily extended into the case with $n$ nodes. Before we start we need one definition. We say a string $x = x_1 x_2...x_n$ is consistent if all substrings $x_i$ are equal. Each processors starts by generating the state
\begin{equation*}
R_i = \frac{1}{\sqrt{2}}(\ket{0}+\ket{1}).
\end{equation*}The state of the system is
\begin{equation*}
\prod_{i}R_i =\frac{1}{\sqrt{2^n}} \sum_{i=0}^{2^n-1}\ket{i}.
\end{equation*}
Each processor stores the consistency of the system in the qubit $\ket{S_i}$. The circuit for the process is shown in Fig 3

\begin{figure}
\begin{align*}
\includegraphics[width=0.4\textwidth]{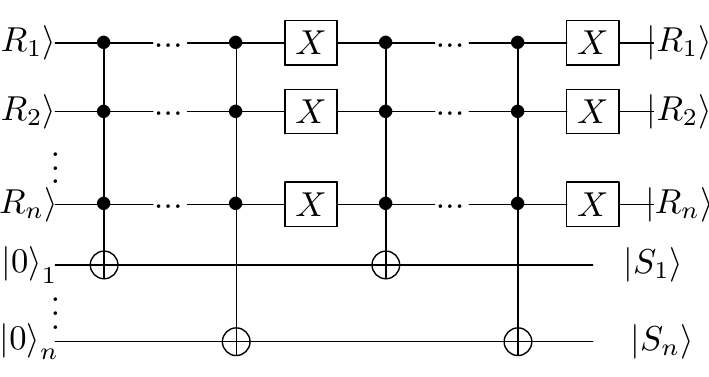}
\end{align*}
\caption{}
\end{figure}
The global system becomes
\begin{align*}
\ket{R_1...R_nS_1...S_n} &= (\ket{0^{\otimes^n}}
+\ket{1^{\otimes^n}})\ket{1^{\otimes^n}} \\
&+ \sum_{i=1}^{2^n-2}\ket{i}\ket{0^{\otimes^n}}.
\end{align*}
Each processor now measures its $\ket{S}$ state. The system collapse to either 
\begin{equation*}
 (\ket{0^{\otimes^n}}+\ket{1^{\otimes^n}})\ket{1^{\otimes^n}} 
\end{equation*}
or 
\begin{equation*}
 \sum_{i=1}^{2^n-2}\ket{i}\ket{0^{\otimes^n}}.
\end{equation*}
If the system collapses to 
\begin{equation*}
\sum_{i=1}^{2^n-2}\ket{i}\ket{0^{\otimes^n}}
\end{equation*}
then the system $R_1$...$R_n$ is inconsistent. Since the system is inconsistent at least one $R_i = \ket{0}$ and at least one $\ket{R_j} = \ket{1}$. Any processor that measures its $\ket{R_i }= \ket{1}$ is a leader candidate. Any processor that measures its $\ket{R_i }= \ket{0}$ is a follower. There are now at most $(n-1)$ leader candidate, for which the process is repeated. 
In the event that the system collapsed to the $( \ket{0^{\otimes^n}}+\ket{1^{\otimes^n}})\ket{1^{\otimes^n}} $ state, we need to apply a unitary such that the symmetry is broken. If the number of states $n$ is even then we apply the unitary 
\begin{equation*}
U = \frac{1}{\sqrt{2}} 
\begin{pmatrix}
1 & e^{-i\pi/n} \\
-e^{i\pi/n} & 1
\end{pmatrix},
\end{equation*}
to each $\ket{R_i}$, which we will show breaks the symmetry. For symmetry to be preserved the system $\ket{R_1...R_n}$ must be in the state $\ket{0}^{\otimes^n}$ or $\ket{1}^{\otimes^n}$.  After each processor applies $U$ to $\ket{R_i}$, the probability of being in either of these states is 
\begin{align*}
\text{Prob}(\ket{0}^{\otimes^n}) 
&= \frac{1}{\sqrt{2}} \Big[ \Big(\frac{1}{\sqrt{ 2}}\Big)^n +\Big(\frac{e^{i\frac{\pi}{n}}}{\sqrt{ 2}}\Big)^n  \Big],\\
&=0,
\end{align*}
and
\begin{align*}
\text{Prob}(\ket{1}^{\otimes^n})& = \frac{1}{\sqrt{2}} \Big[ \Big(\frac{1}{\sqrt{ 2}}\Big)^n +\Big(\frac{-e^{i\frac{\pi}{n}}}{\sqrt{ 2}}\Big)^n  \Big],\\
&=0.
\end{align*}
Thus, after applying $U$ the probability of being in a symmetric state is zero. After applying $U$, if a processor measures its $\ket{R_i} $ to be $\ket{1}$, it is a leader candidate. Since the system is in an asymmetric state, at least one processor will lose eligibility, and at least one processor will remain eligible. 
If the number of states $n$ is odd we cannot simply apply $U$. Instead, for each processor we need an additional register $\ket{T_i}$ initialized to $\ket{0}$. Set $T_i = R_i\oplus T_i$. Then apply $V_n$ to $R_i \otimes T_i$. We define $\frac{1}{\sqrt{R_n+1}} V_n$ as the matrix

\begin{equation*}
\begin{pmatrix}
1/\sqrt{2}   & 0      & \sqrt{R_n}                         & e^{i\frac{\pi}{n}}/\sqrt{2} \\
1/\sqrt{2}   & 0      & -\sqrt{R_n}e^{-i\frac{\pi}{n}}     & e^{-i\frac{\pi}{n}}/\sqrt{2} \\
\sqrt{R_n}  & 0     & \frac{-i e^{-i\frac{ \pi I_n}{2n}}}{\sqrt{2}R_{2n}}  & -\sqrt{R_n} \\
0                   & \sqrt{R_n +1} & 0                         & 0
\end{pmatrix},
\end{equation*}

where $R_n$ and $I_n$ are the real and imaginary parts of $e^{i\frac{\pi}{n}}$, respectively. This matrix is well defined since $0<|R_n| < 1$.  With some calculations $V_n$ is shown to be unitary. Similar to the case where $n$ is even, for symmetry to be preserved the system must be in one of the following states $\ket{00}^{\otimes^k},\ket{01}^{\otimes^k},\ket{10}^{\otimes^k},\ket{11}^{\otimes^k}$. However, after each processor applies $V_n$, the probability of the system being in any one of these states is
\begin{align*}
\text{Prob}(\ket{00}^{\otimes^n}) &= \frac{1}{\sqrt{2}} \Big[ \Big(\frac{1}{\sqrt{2R_n + 2}}\Big)^n \\
&+\Big(\frac{e^{i\frac{\pi}{n}}}{\sqrt{2R_n + 2}}\Big)^n  \Big],\\
&=0,
\end{align*}
\begin{align*}
\text{Prob}(\ket{01}^{\otimes^n}) &= \frac{1}{\sqrt{2}} \Big[ \Big(\frac{1}{\sqrt{2R_n + 2}}\Big)^n \\
&+\Big(\frac{-e^{i\frac{\pi}{n}}}{\sqrt{2R_n + 2}}\Big)^n  \Big],\\
=0,
\end{align*},\begin{align*}
\text{Prob}(\ket{10}^{\otimes^n}) &= \frac{1}{\sqrt{2}} \Big[ \Big(-\frac{1}{\sqrt{2R_n + 2}}\Big)^n\\
&+\Big(\frac{1}{\sqrt{2R_n + 2}}\Big)^n  \Big] ,\\
&=0,
\end{align*}

\begin{align*}
\text{Prob}(\ket{11}^{\otimes^n}) = 0.
\end{align*}
Thus, the symmetry is broken. Each processor now measures $\ket{R_i\:T_i}$, and the processors with the largest value of $\ket{R_iT_i}$ are candidates for the next round. Again, since symmetry was broken, at least one processor will lose eligibility, and at least one processor will remain eligible.

 \section*{Quantum Consensus and Non-Bias Leader Election}
Up to this point, we have been assuming no faulty processors. Maor Ganz considers the case with a group of $n$ processors who do not trust each other and want to elect a leader. In his paper,  [\newcite{Ganz}], Ganz considers an algorithm that gives an honest processor at least $\frac{1}{n}-\epsilon$ probability of winning. Using classical consensus, this problem was shown to be impossible by Mochon in [\newcite{Mochon}]. However, using quantum consensus Mochon showed that in certain cases one can formulate an algorithm with arbitrarily small $\epsilon$. 

This algorithm is based on a series of quantum coin flips in tournament style. In other words, processors are paired and a single quantum coin flip is used to eliminate a processor from each pair. The main difficulty is in creating fault-tolerant coin-flipping. 

There are two types of bias coin flipping: strong and weak coin-flipping. A strong coin-flipping protocol with bias $\epsilon$ is a protocol in which neither party is capable of forcing the probability of any given flip to be greater than $1/2+\epsilon$. In weak coin flipping, both parties, Alice and Bob, have a predetermined desired coin outcome. For example, a 1 can be thought of as Alice winning and a 0 can be thought of as Bob winning. In weak coin flipping, neither player can shift the probability of the coin flip towards their desired outcome with probability greater than $1/2+\epsilon$. In the classical scenario, weak and strong coin flipping are essentially equivalent. However, as Mochon states "in the quantum world the two are very different," [\newcite{Mochon}]. For this paper, we will only be concerned with weak coin-flipping. 

Machon's algorithm, or rather his proof of the existence of an algorithm for weak coin flipping with arbitrarily small bias is a significant result in quantum algorithms. However, as stated by Ganz "the result has not been peer-reviewed, its novel techniques (and in particular Kitaev’s point game formalism) have not been applied anywhere else, and an explicit protocol is missing," [\newcite{Ganz}]. With that said, the basic setup for weak coin flipping is quite similar to the setup for algorithms II. Figure 4 illustrates the process. Alice starts with state $\ket{\psi_{A,0}}$ on space $A$ and Bob starts with state $\ket{\psi_{B,0}}$ on state $B$. On every odd round Alice applies a unitary $U_{A,i}$ and projection $E_{A,i}$ to space $A\otimes M$, and on every even round Bob applies a unitary $U_{b,i}$ and projection $E_{B,i}$ to space $M\otimes B$. The basic idea is that by applying specific unitaries and projections, an honest player can decrease any bias to arbitrarily small values.

\begin{figure}
\begin{align*}
\includegraphics[width=0.5\textwidth]{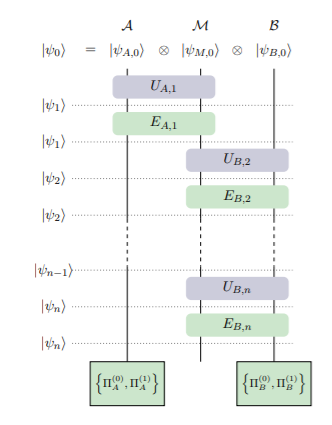}
\end{align*}
\end{figure}
\begin{centering}
Figure 4: retrieved from [\newcite{Aharonov}]
\end{centering}
\\
\\
Mochon's paper proving this result is 80 pages, and we do not have the time to go into detail. However, this is an impressive result in quantum information and demonstrates some of the beauty of the field.

 \section*{Implementation}
By using several existing quantum software packages, we were able to simulate Quantum Leader Election Algorithm II. We used the packages listed below.
\begin{itemize}
\item Qiskit is an open-source SDK for working with quantum computers at the level of pulses, circuits, and application modules. 
\item ProjectQ is an open-source software framework for quantum computing. It provides tools for implementing and running quantum algorithms using either classical hardware. 
\item SimulaQron allows distributed simulation of the nodes in a quantum internet network.
\end{itemize}
The source of implementation for simulating Quantum Leader Election Algorithm can be found at \url{https://github.com/lifuzhang1108/ quantum-consensus}.

\textbf{Pseudo code:}

\begin{enumerate}
\item Prepare one-qubit quantum registers $R_1$,...,$R_6$, $T_1,...,T_6$ and $S_6,...,S_6$. 

\item For each processor, do the following: 
 
\item If $status$ = “eligible,”\\
Set $\ket{R_i} = (\ket{0}+ \ket{1})/ \sqrt{2}$ \\
Set $\ket{S}=\ket{0}$;\\

\item Apply circuit in figure 3.

\item Measure $\ket{S}$. \\

\item If $\ket{S}=\ket{0}$, measure $\ket{R_i}$.\\
If $\ket{R_i} = \ket{1}$, $status$ = “eligible.” \\
If $\ket{R_i} = \ket{0}$, $status$ = “ineligible.” \\ 
    
\item If $\ket{S}=\ket{1}$\\
and there are an even number of eligible processors,\\
apply unitary $U$ to $\ket{R_i}$,\\
measure $\ket{R_i}$.\\ 
If $\ket{R_i} = \ket{1}$, $status$ = “eligible,” \\
If $\ket{R_i} = \ket{0}$, $status$ = “ineligible,” \\ 
    
\item  If $\ket{S}=\ket{1}$\\
 and there are an odd number of eligible processors,\\
initialize $T_i$ and apply unitary $V_n$ to $\ket{R_i T_i}$.\\
Measure all $\ket{R_i T_i}$ 
If $R_i T_i = max(R_1 T_1,..,R_n T_n)$, $status$ = “eligible,” \\
If $R_i T_i < max(R_1 T_1,..,R_n T_n)$, $status$ = “ineligible,” \\

\item Output status.
\end{enumerate}
 
 \section*{Summary}
Quantum computing provides tools for achieving consensus in a distributed system. It is shown by Tani that the classically non-deterministic anonymous leader election problem is can be solved deterministically using quantum computers.  As a proof of concept demonstration, we implemented the quantum algorithms in Qiskit and simulated quantum network using SimulaQron, the algorithm can successfully elect a single leader among anonymous parties. Mochon demonstrated that quantum consensus algorithms can be used to fairly elect a leader even under Byzantine conditions. While these algorithms have not been used in practice, they offer excellent insight into both information theory and quantum mechanics.

\bibliography{nejlt}
\bibliographystyle{nejlt_bib}

\end{document}